\documentclass[twocolumn,nofootinbib]{revtex4}

\usepackage[T1]{fontenc}
\usepackage{amsmath}
\usepackage{esint}
\usepackage{hyperref}
\usepackage{color}
\usepackage[normalem]{ulem}
\makeatletter

\newcommand*\xbar[1]{%
  \hbox{%
    \vbox{%
      \hrule height 0.5pt 
      \kern0.3ex
      \hbox{%
        \kern-0.0em
        \ensuremath{#1}%
        \kern-0.0em
      }%
    }%
  }%
}

\usepackage{amsfonts}

\usepackage{bbm}

\newcommand{\be}{\begin{equation}}
\newcommand{\ee}{\end{equation}}
\newcommand{\bea}{\begin{eqnarray}}
\newcommand{\eea}{\end{eqnarray}}

\begin{document}

\title{\boldmath The Bondi-Metzner-Sachs group in five spacetime dimensions}

 \author{Oscar Fuentealba$^{1}$} 
\author{Marc Henneaux$^{1,2}$} 
\author{Javier Matulich$^{1}$}
\author{C\'{e}dric Troessaert$^{3}$}

\affiliation{$^{1}$Universit\'e Libre de Bruxelles and International Solvay Institutes, ULB-Campus Plaine CP231, B-1050 Brussels, Belgium}
\affiliation{$^{2}$Coll\`ege de France, 11 place Marcelin Berthelot, 75005 Paris, France}
\affiliation{$^{3}$Haute-Ecole Robert Schuman, Rue Fontaine aux M\^{u}res, 13b, B-6800, Belgium}



\begin{abstract}
{\bf Abstract:}
We study asymptotically flat spacetimes in five spacetime dimensions by Hamiltonian methods, focusing on spatial infinity and keeping all asymptotically relevant nonlinearities in the transformation laws and in the charge-generators. Precise boundary conditions that lead to a consistent variational principle are given.  We show that the algebra of asymptotic symmetries, which had not been uncovered before, is a nonlinear deformation of the semi-direct product of the Lorentz algebra by an abelian algebra involving four independent  (and not just one) arbitrary functions of the angles on the $3$-sphere at infinity, with non trivial central charges. The nonlinearities occur in the Poisson brackets of the boost generators with themselves and with the other generators. They would be invisible in a linearized treatment of infinity.  

\end{abstract}

\maketitle

The structure of null infinity in five (more generally in odd) spacetime dimensions is rather complicated \cite{Hollands:2003ie,Hollands:2003xp,Hollands:2004ac,Hollands:2016oma}.  This is because the gravitational field decays with negative fractional powers of $r$ as one tends to infinity along null curves, preventing a smooth conformal compactification of spacetime along the lines proposed by Penrose \cite{Penrose:1962ij}.  

By contrast, there is no particular feature that distinguishes even and odd spacetime dimensions at spatial infinity.  The gravitational field decays with a Coulomb type $\sim r^{-D+3}$ behaviour up to diffeomorphisms, where $D$ is the number of spatial dimensions. This fall-off allows gravitational radiation with finite energy, which ``has not arrived yet (and will never arrive) at spatial infinity''.

What makes spatial infinity simpler from this point of view is that its existence is not a dynamical question,  contrary to the existence of a null infinity with given smoothness properties, which is a delicate dynamical question even in four spacetime dimensions \cite{Friedrich2004,ValienteKroon:2002fa}.  How a simple asymptotic behaviour of the fields at spatial infinity leads to a more subtle behaviour at null infinity as one follows initial data by integrating the dynamical equations  is eloquently illustrated in the study of the Maxwell field in five spacetime dimensions, where the radiative branch develops a $r^{-\frac32}$ behaviour, while the Coulomb branch remains of order $r^{-2}$ \cite{Henneaux:2019yqq}.

It was long thought that the Bondi-Metzner-Sachs (BMS) group in four spacetime dimensions, initially discovered at null infinity \cite{Bondi:1962px,Sachs:1962wk,Sachs:1962zza},  could not be seen at spatial infinity.  
This negative conclusion was recently realized to be due to a too restrictive choice of boundary conditions \cite{Henneaux:2018cst,Henneaux:2018hdj,Henneaux:2019yax}.  In the approach developed in  \cite{Henneaux:2018hdj,Henneaux:2019yax}, boundary conditions that differ by an improper gauge transformation \cite{Benguria:1976in} from the boundary conditions of  \cite{Regge:1974zd} were proved to lead to the full BMS$_4$ algebra.   
Improper gauge fixing  (i.e., fixing of improper gauge transformations or `large diffeomorphisms') might  (and in fact, does) hide important physical information.

The study of the asymptotic structure of gravity at spatial infinity in five spacetime dimensions is a conceptually direct extension of the four-dimensional analysis of \cite{Henneaux:2018hdj,Henneaux:2019yax}.  It reveals, however, interesting new features not uncovered before. 

First of all, the relevant BMS$_5$ group involves supertranslations parametrized by four independent functions of the angles on the sphere at infinity, instead of just one as a naive generalization from four dimensions might have led one to anticipate \cite{Tanabe:2009xb}. 

Second, the algebra of the symmetry generators is nonlinear.  There is in fact nothing surprising in the property that symmetry generators form a nonlinear algebra as discussed in the lucid review \cite{deBoer:1995cqx}. It is a familiar phenomenon  in classical mechanics. Asymptotic symmetries are no exception in that respect.  The first examples were found in three dimensions \cite{Henneaux:1999ib,Henneaux:2010xg,Campoleoni:2010zq}, and more recently, in the case of supergravity in four dimensions \cite{Fuentealba:2021xhn}.  That the asymptotic symmetry algebra of gravity in higher spacetime dimensions is nonlinear is a new result which is not in contradiction with the linear findings of null infinity studies, which restricted the asymptotic analysis to linear terms and were thus blind to the nonlinearity exhibited here.

The Hamiltonian action of General Relativity in five spacetime dimensions reads
\begin{eqnarray}
&& \hspace{-.4cm}S[g_{ij},\pi^{ij},N,N^i]= \nonumber \\
&&  \hspace{-.4cm} \int dt \left[\int d^{4}x \left(\pi^{ij}\dot{g}_{ij}-N \mathcal{H}- N^{i} \mathcal{H}_{i}\right)-B_{\infty}\right]\,. \hspace{.4cm} 
\end{eqnarray}
Here, $\pi^{ij}$ corresponds to the conjugate momentum of the four-dimensional spatial metric $g_{ij}$, while $N$ and $N^i$ stand for the lapse and shift functions, respectively, which we take to behave asymptotically as $N\rightarrow 1,\, N^i \rightarrow 0$. The surface integral on the $3$-sphere at spatial infinity $B_\infty$ coincides with the standard ADM energy with this asymptotic behaviour of the lapse and the shift. 

Variation with respect to the Lagrange multipliers $N$ and $N^i$ implies the "Hamiltonian" and "momentum" constraints
$\mathcal{H}  =\frac{1}{\sqrt{g}}\left(\pi^{ij}\pi_{ij}-\frac{\pi^{2}}{3}\right)-\sqrt{g}R\approx 0$ and $\mathcal{H}_{i}  =-2\nabla^{j}\pi_{ij}\approx 0$, respectively. 

The usual fall-off the fields in four spatial dimensions is given by (see e.g. \cite{Jamsin:2007qh,Jamsin:2008dza}) $g_{ij}=\delta_{ij}+\overline{h}_{ij}r^{-2}+h^{(2)}_{ij}r^{-3}+\mathcal{O}\left(r^{-4}\right)$ and $ \pi^{ij}=\overline{\pi}^{ij}r^{-3}+\pi^{(2)ij}r^{-4}+\mathcal{O}\left(r^{-5}\right)$.
This decay of the fields is preserved under the action of the Poincar\'e group, where the asymptotic behaviour of the parameters reads
$\xi=b_ix^i+a_0+\mathcal{O}\left(r^{-1}\right)$ and  $
\xi^i=b^i_{\,\,\,j}x^j+a^i_0+\mathcal{O}\left(r^{-1}\right)$.
The constant $b_i$ parametrizes Lorentz boosts, while $b_{ij}=-b_{ji}$ generates spatial rotations. The constants $a_0$ and $a^i_0$ correspond to standard translations.  Note that the term $b^ix^0$ in $\xi^i$ can be absorbed in a spatial translation  $a^i_0$ at any given time.  

With the above fall-off, there is no room for supertranslations since these will induce terms of order $r^{-1}$ in the metric and of order $r^{-2}$ in its conjugate momentum, violating the boundary conditions. However, one can consistently relax these asymptotic conditions in such a way that the asymptotic symmetry is enlarged to include supertranslations. By ``consistently'', we mean that the action -  in particular the kinetic term - and the charges remain finite.
To this end, we follow the prescription introduced in the case of gravity \cite{Henneaux:2018hdj,Henneaux:2019yax} and supergravity \cite{Fuentealba:2021xhn} in four dimensions, and in electromagnetism in five dimensions \cite{Henneaux:2019yqq}, i.e., we allow an improper gauge transformation term in the asymptotic behaviour of the fields.  We also request, as in \cite{Henneaux:2018hdj,Henneaux:2019yax} and for the same reasons, that the improper diffeomorphism (written in Hamiltonian form) preserves the condition $h_{rA} = 0$ to leading $\mathcal{O}(1)$ order. This makes the improper gauge terms in the metric and its momentum depend on two arbitrary functions $U$ and $V$ of the angles, which are not restricted by parity conditions in five spacetime dimensions.

The resulting fall-off in spherical coordinates $(r, x^A)$ is given by
\begin{align}
    g_{rr} &= 1 + \frac {2 \xbar \lambda} {r^2} +\frac {h^{(2)}_{rr}} {r^3} + \mathcal{O}\left(r^{-4}\right)\,, \nonumber\\
    g_{rA} &= \frac{\xbar \lambda_A}{r} + \frac{h^{(2)}_{rA}}{r^2} + \mathcal{O}\left(r^{-3}\right)\,, \\
    g_{AB} &= r^2 \xbar g_{AB} + r \theta_{AB} + \xbar h_{AB} +\frac {h^{(2)}_{AB}} {r} + \mathcal{O}\left(r^{-2}\right)\,, \nonumber 
 \end{align}
 and 
 \begin{align}
    \pi^{rr} &= r \kappa^{rr} + \xbar \pi^{rr} +\frac{\pi^{(2)}_{rr}}{r}+ \mathcal{O}\left(r^{-2}\right)\,, \nonumber \\
    \pi^{rA} &= \kappa^{rA} + \frac{\xbar \pi^{rA}}{r} + \frac{\pi^{(2)rA}}{r^2} + \mathcal{O}\left(r^{-3}\right)\,,  \\
    \pi^{AB} &= \frac{\kappa^{AB}}{r} + \frac{\xbar \pi^{AB}}{r^2}+  \frac{\pi^{(2)AB}}{r^3}+ \mathcal{O}\left(r^{-4}\right)\,. \nonumber
\end{align} 
The functions  $\theta_{AB}$,  $\kappa^{rr}$, $ \kappa^{rA} $ and  $ \kappa^{AB} $ generated by the improper gauge transformation read, in terms of $U$ and $V$, $ \theta_{AB} = \xbar D_A \xbar D_B U + \xbar g_{AB} U$, $\kappa^{rr} = \sqrt {\xbar g} \, \xbar \triangle V$, $\kappa^{rA} = \sqrt {\xbar g} \, \xbar D^A V$ and $\kappa^{AB} = \sqrt {\xbar g} \, (\xbar g^{AB} \xbar \triangle V - \xbar D^A \xbar D^B V)$, where $\xbar \triangle\equiv \xbar D_A \xbar D^A$. Here, $\xbar D_A$ denotes the covariant derivative associated to the unit metric $\xbar g_{AB}$ on the $3$-sphere at spatial infinity. 

For this new set of asymptotic conditions the fall-off (in spherical coordinates) of the constraints is given by
$\mathcal H = \mathcal O\left(r^{-1}\right)$, $\mathcal H_r = \mathcal O\left(r^{-1}\right)$ and 
   $ \mathcal H_A = \mathcal O\left(1\right)$.  Just as in four dimensions \cite{Henneaux:2018hdj,Henneaux:2019yax}, we shall impose that the constraints decay faster at infinity, i.e., we require $\mathcal H = \mathcal O\left(r^{-3}\right)$, 
    $\mathcal H_r = \mathcal O\left(r^{-2}\right)$ and 
    $\mathcal H_A = \mathcal O\left(r^{-1}\right)$.
These conditions do not eliminate physical solutions, since for such solutions the constraints are satisfied to all orders.  We stress that only the constraints are imposed asymptotically, not the dynamical equations.  With these boundary conditions, the kinetic term in the action and hence the symplectic structure
 $\Omega = \int d^4 x\, d_V \pi^{ij} \, d_V g_{ij} $ can be verified to be finite \cite{InPreparation}.

It is worth pointing out that in four spacetime dimensions, the improper gauge terms and the ``core terms'' of the metric are at the same order in $1/r$ but come with distinct parity properties, which enable one to separate them.  In five spacetime dimensions, they correspond to different powers of $1/r$ and so can also be distinguished even though there is no parity condition.

The new set of asymptotic conditions is preserved under asymptotic diffeomorphisms generated by the following parameters
\begin{align}
\xi & =br+T  
+\frac{1}{r} T^{(1)} + \hbox{``more''} + \mathcal{O}(r^{-2})\,,  \nonumber\\
\xi^{r} & =W+\frac{1}{r}W^{(1)}+\mathcal{O}(r^{-2})\,,  \\
\xi^{A} & =Y^{A}+\frac{1}{r}\overline{D}^{A}W 
+\frac{1}{r^{2}}I_{(1)}^{A} + \hbox{``more''} +\mathcal{O}(r^{-3})\,,  \nonumber
\end{align}
where $b= b_i x^i$ stands for the Lorentz boosts, and $Y^A = \frac12 b_{ij} x^i e^{jA} $ are the Killing vectors of spatial rotations ($e^{jA}$ are the vectors tangent to the $3$-sphere defined as in \cite{Henneaux:2019yax}). The functions $T$ and $W$ are arbitrary functions on the $3$-sphere and describe the natural generalization to five dimensions of  four-dimensional supertranslations.  Ordinary time translations correspond to the zero mode $T_0$ while spatial translations correspond to the first spherical harmonics of $W$. The functions $T^{(1)}$, $W^{(1)}$ and $I^A_{(1)}$ associated with the next power of $1/r$ are also arbitrary functions on the $3$-sphere and are kept because they define independent non trivial symmetries with non-vanishing charges.  However, only $T^{(1)}$ and the combination $\tilde{I}^{(1)}=\xbar D_A I^{(1)A}-\xbar \triangle W^{(1)}$  actually appear in the expression of the charges, so that transformations for which $\tilde{I}^{(1)} = 0 $ are proper gauge transformations.  Because the transformations parametrized by $T^{(1)}$ and  $\tilde{I}^{(1)} $ form an abelian algebra with the supertranslations  (centrally extended, see below), they are also called (subleading) ``supertranslations''. Finally,  ``more'' denotes correcting terms which must be included in order to preserve the boundary conditions and make the charges integrable.  Their explicit expression is not particularly illuminating for our purpose so that they will not be reproduced here \cite{Correction terms} (see \cite{InPreparation}).  Let us simply recall that similar terms appear already in four spacetime dimensions \cite{Henneaux:2018cst,Henneaux:2018hdj,Henneaux:2019yax}.

The canonical generators $G_X$ associated with the phase space vectors $X$ defining these asymptotic diffeomorphisms can be obtained through the standard symplectic rule $\iota_X \Omega = - d_V G_X$ \cite{Henneaux:2018gfi}.  In the present case where the symplectic $2$-form $\Omega$ reduces to the standard $dp \wedge dq$ bulk piece, this rule is equivalent to the Regge-Teitelboim condition that the generators $G_X$ should have well-defined functional derivatives  \cite{Regge:1974zd}. 

By applying this rule, the charge-generators are found to be the usual sum of a bulk piece proportional to the constraints and a surface integral over the sphere at infinity.   The surface integrals  are finite (and integrable) thanks to the constraints and take the form
\begin{equation}
Q_{\xi}=b_iB^i+\frac{1}{2}b_{ij}M^{ij}+Q_{T}+Q_{W}+ Q_{T^{(1)}}+Q_{I^{(1)}}\,,  
\end{equation}
with $Q_{T}=\oint d^3x \sqrt{\xbar g }\,T\, \mathcal T $, $ Q_{W}=\oint d^3x \sqrt{\xbar g }\,W\, \mathcal W$ and
\begin{align}
&Q_{T^{(1)}}=-\oint d^3x \sqrt{\xbar g }\, T^{(1)}((\xbar \triangle+3)U)\,, \label{eq:ChargeU}\\
& Q_{I^{(1)}}=-2\oint d^3x \sqrt{\xbar g} \tilde{I}^{(1)}V\,.  
\end{align}
The expression of the Lorentz charges $B^i$ and $M^{ij}$ is rather cumbersome and involves up to cubic terms in the asymptotic fields.  Similarly, the integrands $\mathcal T$ and  $\mathcal W$ turn out to be quadratic.  The corresponding explicit formulas will be given elsewhere \cite{InPreparation}.  Here, we shall focus only on the properties of their Poisson bracket algebra, which exhibits interesting features.  Before doing this, we stress that the asymptotic symmetry algebra is parametrized by the $10$ Lorentz parameters $b_i$ and $b_{ij}$ as well as by four independent functions on the $3$-sphere ($T$, $W$, $T^{(1)}$ and $\tilde{I}^{(1)}$) \cite{Correction terms}. 

The Hamiltonian formulation is particularly transparent for the purpose of computing the commutator  of asymptotic symmetries (which are also asymptotic symmetries with well-defined generators according to general theorems  \cite{Brown:1986ed}) since the symplectic structure is given ab initio from the action without need to guess it.  
Because the symmetry generators are gauge-invariant, i.e., invariant under proper gauge transformations, one can either use the Poisson bracket to compute the algebra (keeping any of the accompanying weakly vanishing bulk piece with the understanding that two symmetry generators that coincide modulo constraints should be identified) or use the Dirac bracket.  In order to avoid gauge fixing, we shall use the first method, but without always writing explicitly the physically irrelevant matching bulk piece. 

One striking feature is that the Lorentz subalgebra, specifically, the bracket of two boosts, contains a term cubic in the charges. Explicitly,
\be
\{B^i,B^j\}=M^{ij}+2\oint d^3x \sqrt{\xbar g}\,x^{[i}e^{j]A} \Lambda_{AB} K^B  
\ee
with $K_A=\xbar D_A V$.  Here, 
\begin{eqnarray}
&&\Lambda_{AB} = 8 K_A K_B-\frac{3}{2}\xbar g_{AB}(\xbar \triangle U+3U)^2 \nonumber \\
&& -2(\xbar D_A \xbar D_BU+\xbar g_{AB}U)(\xbar \triangle U+3U)  
\end{eqnarray}
The other Lorentz commutation relations involving spatial rotations are unchanged (and linear).

The brackets of the Abelian generators $U$ and $V$ with the Lorentz generators are given by
\begin{align}
\{B^i,U\}&=-2n^i V  \,, \\
\{M^{ij},U\}&=-\xbar D_B(x^{[i}e^{j]B} U)\,, \\
\{B^i,V\}&=\frac{1}{2}\Big[4n^i U+\partial_A n^i \xbar D^A U+n^i \xbar \triangle U \Big]\,,   \\
\{M^{ij},V\}&=-\xbar D_B(x^{[i}e^{j]B} V)\,. 
\end{align}
The functions $U$ and $V$ transform therefore in linear representations of the Lorentz algebra (which is consistent, even though the brackets of the boosts are nonlinear).  This representation is the direct analog, in one dimension higher, of the representation of the supertranslations in four spacetime dimensions written in the Hamiltonian basis \cite{Henneaux:2018cst,Troessaert:2017jcm}. Note that the factor $4$ in the bracket $\{B^i,V\}$ was $3$ in the corresponding lower dimensional formula.
 
The brackets of the supertranslations generated by ${\mathcal T}$ and ${\mathcal W}$ with the Lorentz generators describe the same representation, augmented by nonlinear terms in the brackets with the boosts, specifically,
\begin{align}
\{B^i,\mathcal T\}&=-n^i \mathcal W+\Lambda^i_{\mathcal T}\,,  \\
\{B^i,\mathcal W\}&=-4n^i \mathcal T-\partial_A n^i \xbar D^A \mathcal T-n^i \xbar \triangle \mathcal T+\Lambda^i_{\mathcal W}\,,  
\end{align}
The explicit expression of the nonlinear terms $\Lambda^i_{\mathcal T}$ and $\Lambda^i_{\mathcal W}$, which are quadratic, will be given elsewhere \cite{InPreparation}.  The explicit  check that they are compatible with the Jacobi identity is particularly instructive.

Finally, the nonvanishing brackets between supertranslations and the Abelian generators $U$, $V$ are
\begin{align}
\{\mathcal T(x),V(\xbar x)\}&=- \delta^{(3)}(x-\xbar x)\,, \\
\{\mathcal W(x),U(\xbar x)\}&=2\delta^{(3)}(x-\xbar x)\,.  
\end{align}
They involve a central charge.  Note the remarkable similarity with the structure found in the asymptotic treatment of four-dimensional supergravity, where there are also two sets of fermionic charges forming an abelian algebra with a non-trivial central extension \cite{Fuentealba:2021xhn}.  The similarity goes beyond this feature, since there are also nonlinear terms in the algebra and furthermore, the leading improper gauge functions in the expression of the fields (here $U$ and $V$) turn out to be the charges for the symmetry parametrized by the subleading terms in the asymptotic expansion of the gauge parameters.

In this letter, we have investigated the asymptotic structure of Einstein gravity in five spacetime dimensions by focusing on spatial infinity and using  Hamiltonian techniques.  The choice of five dimensions was made not only for reasons of simplicity, but also with the purpose of emphasizing that the discomfort characteristic of null infinity analyses in odd spacetime dimensions is simply not present at spatial infinity.    
We computed the algebra without having to linearize the charges at infinity, exhibiting thereby its interesting nonlinear structure. This is a new algebra to the best of our knowledge. We have not investigated in detail whether the nonlinearities could be removed by nonlinear redefinitions, but we stress that they do occur with the natural Hamiltonian choices of parametrizations of the symmetries made here.

The key to revealing the full symmetry structure at spatial infinity is to include  improper gauge transformation terms in the metric and its conjugate momentum, as in four dimensions \cite{Henneaux:2018hdj,Henneaux:2019yax}.  The inclusion of improper diffeomorphisms has also been advocated at null infinity in four dimensions to account for the subleading soft theorems \cite{Campiglia:2014yka,Campiglia:2015kxa,Campiglia:2016jdj,Campiglia:2016efb}, as well as in higher (mostly even) dimensions \cite{Kapec:2015vwa,Campiglia:2017xkp,Pate:2017fgt,Aggarwal:2018ilg,Campoleoni:2020ejn,Capone:2020mwy,Strominger:2017zoo}, and near the horizon \cite{Grumiller:2019fmp}.  Finally, we expect our results to generalize to higher dimensions in a conceptually straightforward way. It would also be of interest to work out the asymptotic behaviour of the fields as one goes to infinity along null geodesics, along the lines of \cite{Henneaux:2019yqq}.

\section*{Acknowledgments}
Discussions with Sucheta Majumdar and Ricardo Troncoso are gratefully acknowledged. This work was partially supported by the ERC Advanced Grant ``High-Spin-Grav'', by FNRS-Belgium (conventions FRFC PDRT.1025.14 and IISN 4.4503.15), as well as by funds from the Solvay Family.




\begin{thebibliography}{99}
\bibitem{Correction terms} 
  Actually, as Eq.(\ref{eq:ChargeU}) shows, the first spherical harmonics of $T^{(1)}$ define proper gauge transformations with vanishing charges since these are annihilated by the operator $\xbar \triangle+3$, but this removes only a finite number of functions in the infinite-dimensional space of all spherical harmonics.
We will ignore this subtlety here, in particular, in the algebra of the charges given below where the relevant projections implementing this feature are not explicity written. The details will be given in \cite{InPreparation}.
\end{thebibliography}
\end{document}